\documentclass[12pt]{article}
\usepackage[top = 2 cm, bottom = 2 cm, left = 2.5 cm, right = 2 cm]{geometry}
\usepackage{authblk, cite, color, amssymb, amsmath, epsfig, graphicx, graphics}
\usepackage[toc]{appendix}
\usepackage[hypertex, colorlinks = true, linkcolor = blue, citecolor = red]{hyperref}
\usepackage[dvipsnames]{xcolor}

\begin{document}

\title{Can the quasi-molecular mechanism of recombination decrease the Hubble tension?}

\author{{\bf Revaz Beradze} and {\bf Merab Gogberashvili}}
\affil{\small $^1$ Javakhishvili Tbilisi State University, 3 Chavchavadze Avenue, Tbilisi 0179, Georgia \authorcr
$^2$ Andronikashvili Institute of Physics, 6 Tamarashvili Street, Tbilisi 0177, Georgia}

\maketitle

\begin{abstract}

In the recently suggested non-standard, quasi-molecular mechanism of recombination, the presence of neighboring proton increases the ionization energy and decreases the final recombination rate of hydrogen. Both these two effects can lead to the larger value of the present expansion rate of the universe obtained using CMB data and standard cosmological model, and thus are able to reduce or resolve the Hubble tension problem. We note also that due to the quasi-molecular channel the recombination began earlier, what potentially can solve the sigma-eight tension, since the CMB-predicted value of the late matter density will be decreased.

\end{abstract}


The remarkable success of the $\Lambda$CDM model is challenged by the discrepancies in the measurements of current Hubble parameter $H_0$, known as the Hubble tension problem. Recent recalibration of cosmic distance ladder with Gaia EDR3 parallaxes and Hubble Space Telescope Photometry of 75 Milky Way Cepheids gives current best estimate \cite{Riess:2020fzl},
\begin{equation} \label{H0L}
H_0 = 73.2 \pm 1.3 ~~ \rm km~sec^{-1}~Mpc^{-1} ~.
\end{equation}

Recently, analysis of gravitationally lensed quasars with measured time delays provided another quite precise independent estimations of $H_0$ from two different experiments, giving $73.3^{+1.7}_{-1.8}$ \cite{Wong:2019kwg} and $74.2^{+2.7}_{-3.0}$ \cite{Shajib:2019toy}, being in a good agreement with previous local measurements. Motivation for precise knowledge of the Hubble constant and the most prominent methods for its measurement have been summarized in \cite{Freedman:2010xv}.

The Hubble parameter also can be measured in earlier cosmological epochs and (assuming the standard $\rm \Lambda CDM$ cosmological model) used to estimate current expansion rate. The most powerful tool for determining cosmological parameters is Cosmic Microwave Background (CMB) data, which estimates a number for the current expansion rate of the universe with an amazing precision \cite{Aghanim:2018eyx},
\begin{equation} \label{H0P}
H_0 = 67.4 \pm 0.5 ~~ \rm km~sec^{-1}~Mpc^{-1}~.
\end{equation}

To explain discrepancy between the values (\ref{H0L}) and (\ref{H0P}) it is important to measure $H_0$ independently of CMB data and the local distance ladder method. Recent intermediate scale gravitational waves observations \cite{Abbott:2019yzh}, provided an estimation $69^{+16}_{-8}$, central value of which agrees well with (\ref{H0P}), but, having a big uncertainty accurately covers local measurements (\ref{H0L}) within $1\sigma$ error bar. Meanwhile, the weakly model-dependent approach \cite{Nunes:2020uex}, based on the analyzes of transversal BAO scale, in combination with BBN and gravitationally lensed quasars information are in accordance with local measurements (\ref{H0L}).

The $\rm CMB + \Lambda CDM$ model value (\ref{H0P}) is in conflict with local measurements ($\ref{H0L}$) with about $5 \% $, nevertheless we note that it is very sensitive to some cosmological parameters. Different theoretical models that might resolve the Hubble tension problem can be found in the recent reviews \cite{Lin:2019htv, Knox:2019rjx, DiValentino:2021izs}. The discrepancy may be due to some unknown systematics \cite{Efstathiou:2013via, Freedman:2017yms, Rameez:2019wdt}, or can be related to screening effect that could be ruining distance ladder calibrations in the local universe \cite{Desmond:2019ygn}, but it can also be hinting on some new physics beyond the $\rm \Lambda CDM$ model. Such scenarios include the hypotheses of decaying dark matter \cite{Berezhiani:2015yta}, missing of some dark sector \cite{Bernal:2016gxb, Karwal:2016vyq, Poulin:2018cxd, Agrawal:2019lmo, Alexander:2019rsc, Lin:2019qug, Choi:2019jck}, alternate dynamics of dark energy \cite{Mortsell:2018mfj, Yang:2018euj, DiValentino:2019ffd}, neutrino effects \cite{Kreisch:2019yzn, Blinov:2019gcj, Sakstein:2019fmf}, emerging spatial curvature \cite{Bolejko:2017fos}, evolving scalar fields \cite{Panpanich:2019fxq, Smith:2019ihp}, primordial non-Gaussianity \cite{Adhikari:2019fvb}, dissipative axion \cite{Berghaus:2019cls}, the lack of knowledge of the path that CMB photons covered since decoupling \cite{Clarkson:2014pda}, {\it etc}. The Hubble tension can be reduced also using modified scenarios of recombination \cite{Chiang:2018xpn}.

Note that there is also the milder tension between the constraints from CMB data and local measurements on the Universe matter density and the amplitude of matter fluctuations \cite{Battye:2014qga}. Matter density perturbation amplitude $\sigma_8$ is extracted from CMB power spectrum analysis \cite{Aghanim:2018eyx} and is also locally measured using large scale structures. Sunyaev-Zeldovich cluster count by {\it{Planck}} \cite{Ade:2015fva} and analysis of cosmic shear measurement data from several collaborations \cite{Abbott:2017wau} give $> 2\sigma$ lower central values of $\sigma_8$ and matter density then the numbers obtained by CMB power spectrum.

According to standard recombination scenario \cite{Sunyaev:2009qn, Galli:2012rf}, primeval plasma consisted of protons (hydrogen ions), helium ions (with $\sim 24 \%$ of total mass of baryonic matter), electrons and photons. At $z \simeq 5000 - 8000$ doubly ionized helium started to recombine, becoming singly ionized, and then at $z \simeq 1600 - 3500$ neutral helium atoms were formed. At $z \simeq 500 - 2000$, when energy of photons dropped below the hydrogen ionization energy, recombination of hydrogen took place and photons decoupled from matter creating CMB.

The major mechanism of radiative recombination \cite{Peebles:1968ja, Zeldovich:1969en} states that proton and electron can form hydrogen atom only in excited state, accompanied by emission of a photon. Direct recombination in the ground state is inefficient as produced radiation has high energy and will ionize neighboring atom, giving no net result. Atoms in a highly excited state cascades down very quickly to the first excited state with principal quantum number $n = 2$. After that, electrons can reach ground state by radiative decay from a $2p$ state by emitting a Lyman-$\alpha$ photon, or decay from a $2s$ state emitting two photons.

Understanding recombination physics properly allows us to calculate $H_0$ accurately. However, the recombination process is not as trivial as discussed above. In order to get a complete picture, many correction terms must be incorporated \cite{Chluba:2010ca}. The Hubble constant is related to a size of sound horizon at recombination $r_*$. An angular sound horizon
\begin{equation} \label{theta}
\theta = \frac {r_*}{d_A(z_*)}~,
\end{equation}
where $d_A(z_*)$ is an angular diameter distance to the last scattering surface (with $z_*$ being a recombination redshift), is precisely measured by the CMB data analysis. With tight constraint on $\theta$, small variation of $d_A(z_*)$ strongly affects the Hubble constant. It was shown that $1 \% $ increase in $d_A(z_*)$ can lead to $5 \% $ rise of $H_0$, that can remove tension with local measurements completely \cite{Clarkson:2014pda}.

Recently, the dependence of the Hubble constant on different recombination parameters was systematically investigated \cite{Liu:2019awo}. The authors had modified a publicly available standard RECFAST code \cite{Seager:1999bc, Wong:2007ym}, adding scaling factors to several atomic parameters. As a result, they found that $H_0$ (which depends on the hydrogen recombination redshift $z_*$) is the most sensitive to the hydrogen ionization energy and to the hydrogen $2s \rightarrow 1s $ two-photon transition rate:
\begin{itemize}
\item If the ionization energy of hydrogen was higher, recombination process would have started at higher temperatures, i.e. at higher $z$, that means, last scattering surface is located further than we expect and thus, the value of the Hubble constant today must be bigger;
\item If $2s \rightarrow 1s $ transition rate was lower, recombination proceeds slower, which is equivalent to the local increment of temperature and $z_*$ is increased again.
\end{itemize}
Both these parameters are atomic constants well-determined by quantum mechanics. Therefore, we can only speak about effective parameters, which can, in principle, alter the recombination picture and decrease the Hubble tension.

In present article we want to estimate the influence of the new non-standard, quasi-molecular recombination mechanism (QMR) \cite{Kereselidze:2019, Kereselidze:2020ljd, Kereselidze:2021wzj} on the values of $H_0$ and $\sigma_8$ obtained from the CMB data.

In pre-recombination period of hydrogen (at $z \gtrsim 2000$), when the temperature and the density of protons were higher than subsequently, the average distance between protons $R$ was comparable with the radius of hydrogen atom in highly excited states $r_n = 2n^2$ \cite{Kereselidze:2019}. Then an electron in primordial plasma was able to bind two protons and form a temporary $2p-1e$ quasi-molecule state (hydrogen molecular ion $H_2^+$) with ionization energy higher than that of hydrogen \cite{Landau}.

The binding energy of an electron in the isolated hydrogen atom is 13.6~eV. The appearance of a second proton increases or decreases the electron binding energy and the ionization energy of electrons in $H_2^+$ for large $n$-s can be estimated as \cite{Kereselidze:2020ljd}
\begin{equation} \label{E}
E \approx \left(\frac {1}{r_n} \pm \frac {3n^2}{2R^2}\right)13.6~eV \approx \left( 1 \pm \frac{3}{16} \right) \frac {13.6}{2n^2}~eV ~,
\end{equation}
where $R \approx 2r_n \approx 4n^2$ is the average distance between protons in the pre-recombination epoch. This almost $20\%$ increase of the electron binding energy for each $n$ leads to the appearance of indirect channels of radiative transitions, which are forbidden in the standard scenario of recombination on an isolated proton \cite{Kereselidze:2019, Kereselidze:2020ljd}.

According to \cite{Kereselidze:2019}, after the emission of a photon, $H_2^+$ is formed in a highly excited repulsive or attractive state. $H_2^+$ in the repulsive state rapidly dissociates into excited hydrogen atom and proton,
\begin{equation} \label{rep}
H_2^+ \rightarrow H^* + p~.
\end{equation}
$H_2^+$ in the attractive state will cascade down to lower states,
\begin{equation} \label{atr}
H_2^+(n) \rightarrow H_2^+ (n-1) + \hbar \omega~,
\end{equation}
and at some point, it may pass into the lower repulsive state and then dissociate according to (\ref{rep}). The probabilities for formation of the hydrogen molecular ion ($H_2^+$) and neutral hydrogen ($H$) in their ground states were calculated to be comparable with each other in this scenario \cite{Kereselidze:2020ljd}.

Here we have to note that there is a tight constraint on the abundance of molecular hydrogen in the Universe, fractional abundance of $H_2$ relative to the total number of baryons is estimated not to exceed $10^{-12}$ during the recombination epoch \cite{Galli:2012rf}. However, quasi-molecular mechanism concerns the period with high temperatures $T \gtrsim 5,000~K$, when no hydrogen atoms exist yet and the ionized quasi-molecules $H_2^+$ are formed temporarily, rapidly dissociating due to scattering events in hot plasma, giving no neutral $H_2$ as a final product. So, QMR can act as a catalyst for the hydrogen recombination, but it does not increase the abundance of molecular hydrogen at later stages of the Universe's evolution.

Since the average distance between protons at $z \sim 2000$ was comparable with the radius of hydrogen molecular ions, that have relatively high ionization energy,
\begin{equation} \label{E-H2}
E_{H_2^+} \approx 16~ {\rm eV}~ (1.87 \times 10^5~ K)~,
\end{equation}
the $H_2^+$ abundance for that temperatures
\begin{equation} \label{T}
T = 2.73 (1+z) ~K \sim 5,000~K~,
\end{equation}
can be high. Also the lifetime of these excited electronic states was about $10^{-9} - 10^{-7}~s$, which is much greater than the duration of collisions \cite{Kereselidze:2020ljd}.

To estimate the order of corrections induced from the early QMR at the temperature (\ref{T}), let us write the Saha equation for hydrogen recombination,
\begin{equation} \label{Saha}
\frac {x^2_e}{1 - x_e} = 2.9 \times 10^{23}{T^{-3/2}}e^{- E/T}~, \qquad (0 < x_e < 1)
\end{equation}
where
\begin{equation} \label{xe}
x_e = \frac {n_e}{n_{H(tot)}} = \frac {n_e}{4.2 \times 10^5 \, \Omega_b h^2} T^{-3}
\end{equation}
is the ionization fraction -- the number density of electrons, $n_e$, per total number of hydrogen nuclei, $n_{H(tot)}$. The baryon number density
\begin{equation}
\Omega_b h^2 \approx 0.02
\end{equation}
is estimated from the CMB data analyses. In the Saha equation (\ref{Saha}), instead of the hydrogen ionization energy
\begin{equation}
E_H = 13.6~ {\rm eV}~ (1.58 \times 10^5~ K)~,
\end{equation}
we should use an effective value, the $H_2^+$ ionization energy (\ref{E-H2}). Then, at $z \approx 2000$ the Saha equation (\ref{Saha}) gives
\begin{equation} \label{}
\frac {x^2_e}{1 - x_e} \approx 50~.
\end{equation}
From this equation we found that
\begin{equation}
x_e \approx 0.98~, \qquad (T \sim 5,000~K)
\end{equation}
i.e. up to $2\%$ of protons had already formed the neutral hydrogen.

\begin{figure}
\begin{center}
\includegraphics[scale=1]{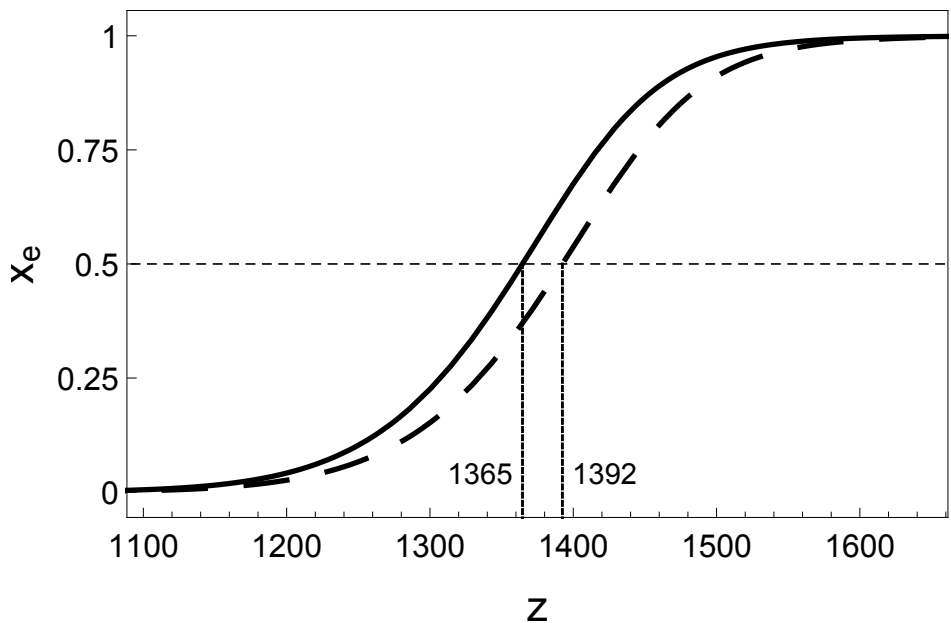}
\caption{Free electron fraction $x_e$ as a function of the redshift $z$. Solid curve shows the standard scenario, while the dashed line corresponds to the modified initial conditions, when the number of free electrons is lower by $2\%$ and the temperature is decreased.}
\label{plot}
\end{center}
\end{figure}

This means that, for the time when the average distance between protons in (\ref{E}) was large, $R \gg r_n$, the quasi-molecular mechanism was in action and by the time the standard mechanism of recombination stepped in, the QMR had prepared different initial conditions:
\begin{itemize}
\item The number of free electrons in (\ref{xe}) was lower by $ \sim 2 \%$ compared to the standard picture;
\item As some photons of plasma had already turned into background radiation, they ceased to contribute to the pressure and the effective temperature in (\ref{Saha}) decreased.
\end{itemize}
If we compare the evolution of the free electron fraction (\ref{xe}) of the standard recombination scenario with the case with modified initial conditions (Figure~\ref{plot}), we see that half of the protons have recombined earlier than expected. More precisely, $x_e=0.5$ at $z=1392$, while the standard treatment of the Saha equation predicted that half of the electrons became bound at $z=1365$, giving a $2\%$ difference.

Of course, it is only qualitative analysis. In a real physical picture, one should consider reaction kinematics as well. In Peebles' three-level atom approximation model \cite{Peebles:1968ja}, free electron fraction evolution differs from the solid curve shown on the figure (\ref{plot}). However, one can argue that the difference between standard and quasi-molecular recombination presented in the simplified model (Figure~\ref{plot}), will persist in more complex physical scenarios.

Now we state:
\begin{itemize}
\item The existence of $H_2^+$ states means that hydrogen recombination processes in the Universe had started earlier ($z \simeq 2000-8000$), when the temperature was greater than required for the standard scenario;
\item Due to the new channel (induced by QMR) for hydrogen formation, the recombination process stretches compared to the case of standard recombination on an isolated proton.
\end{itemize}
The first of these claims is effectively equivalent to the increase in the ionization energy of hydrogen, while the second point can play a role similar to the decrease in $2s \rightarrow 1s$ transition rate. Both these two effects induced from QMR, effectively can act as a rescaling of atomic parameters: hydrogen ionization energy and $2s \rightarrow 1s$ transition rate, which are the most efficient mechanisms to increase the calculated value of $H_0$, according to the analysis in \cite{Liu:2019awo}. It is important also that both these facts have additive effects in estimations of $H_0$.

Noting in addition the quadratic contribution of the Hubble constant into the equation of free electron fraction (\ref{xe}), and accounting the result of \cite{Clarkson:2014pda} that $1\%$ increase in distance to the last scattering surface can give a $5\%$ rise of the $H_0$, we can argue that the $2\%$ inaccuracy in evaluation of the free electron fraction can lead to up to $\sim 4-5\%$ error in CMB estimations of the Hubble constant. Actually, the beginning of hydrogen recombination earlier than expected in the standard picture (due to QMR), can mean that the distance to the last scattering surface is being underestimated in typical analyses.

Another possible consequence of the corrections induced from the QMR can lead to the solution of so-called $\sigma_8$ tension. Note that most attempts at solving the Hubble tension worsen the $\sigma_8$ tension and vice-versa \cite{Hill:2020osr, Jedamzik:2020zmd}. Solutions to the Hubble tension either reduce the size of the sound horizon, $r_*$, or increase the angular diameter distance to the CMB, $d_A$. To keep the locations of the peaks (\ref{theta}) in the CMB fixed, $H_0$ increases, diminishing the tension. On the other hand, a solution to the $\sigma_8$ tension would require either late-universe physics that leads to a suppression of the linear matter power spectrum, or a decrease in the CMB-predicted value of matter density $\Omega_M$. We argue that, if the effects of QMR are important and recombination starts earlier than expected, the CMB analysis may be overestimating the value of $\Omega_M$. Also, since the recombination in QMR scenario lasts longer and ends later, actual linear growth factor can be smaller (in the matter dominated regime the linear growth factor is proportional to the scale factor) that reduces the $\sigma_8$ tension.

A full investigation of cosmological recombination requires the knowledge of wavefunction of an electron in initial and final states. So, it is not so straightforward to carry quantitative analyses as many computational codes need to be modified and it can affect other cosmological parameters, as many of them are correlated. Such calculations are beyond the scope of our short paper. Our analysis of the contribution from $H_2^+$ quasi-molecules into the recombination process, cannot conclusively demonstrate the significance of this phenomena in the Hubble constant and $\sigma_8$ estimations. However, it is clear that the model of recombination may be missing a potentially important ingredient. The QMR, even assuming strong photo-dissociation and secondary ionization processes in hot plasma ($T \gtrsim 5,000~K$), introduces early hydrogen recombination channels and is able to make notable contributions in a complete treatment of the cosmological recombination. In the recent paper about QMR \cite{Kereselidze:2021wzj}, the authors derived algebraic forms of wavefunctions and a scheme of calculations, that in future can allow elaboration of a complete computational code for quasi-molecular mechanism of recombination. The point of our work is to suggest that inclusion of possible corrections due to quasi-molecular processes, along with more well-known mechanisms, in existing computational codes of recombination are important as they may have impact on some cosmological parameters. We tried to demonstrate that inclusion of QMR may affect the estimations of Hubble constant and $\sigma_8$ values from the CMB data analysis in the way to decrease their tensions with local measurements.


\section*{Acknowledgements}

This work was supported by Shota Rustaveli National Science Foundation of Georgia (SRNSFG) through the grant DI-18-335.


\end{document}